\documentstyle[aps,preprint]{revtex}
\begin{document}

\tighten

\title{Heavy quark potential and the phase transitions
in the continuum theory at finite temperature.}

\author{Juraj Boh\'a\v cik}
\address{Institute of Physics SAS, D\'ubravsk\'a cesta 9,
842 28 Bratislava, Slovakia.}
\author{Peter Pre\v snajder}
\address{Dept. of Theoretical Physics, Faculty of Mathematics and Physics,
Comenius Univ., Mlynsk\'a dolina F2, 842 15 Bratislava, Slovakia.}
\maketitle
\vspace{0.5cm}
 \begin{abstract}
Heavy quark potential in the continuum theory at finite temperature
is calculated in different phases by using the Polyakov loop as the order
parameter. We find the linearly rising potential in the confinement phase,
the Debye screened potential in the deconfinement phase and the perturbative
$r^{-2}$ dependence at very high temperatures. Within the approximation
used in this paper we report an evidence of the first order phase transition
accompanied by the $SU(2)$ symmetry restoration at very high temperatures
in the static three dimensional theory.
\end{abstract}
\vspace{1cm}

1. Field theories with one compact dimension
can be used as a tool for the study of interesting statistical
features of the renormalizable four dimensional theories.
In the case of the Euclidean theories with the	compactified imaginary time,
they are known as the finite temperature theories.
The Fourier
expansion in the compactified dimension enables us to reduce the
four dimensional theory to an effective three dimensional field theoretical
model. The Fourier field components acquire the frequency dependent masses,
and therefore it looks reasonable to apply
the Appelquist and Carazzone decoupling theorem\cite{cara}
to the nonzero frequency modes
to calculate an effective theory for static field modes. This process,
generally known as dimensional reduction, was described by Appelquist and
Pisarski\cite{appis} as a possible calculation scheme for high temperature
behaviour of the field theories. However, Landsman\cite{land} proved
in the framework of the finite-temperature renormalization group formalism
why QCD does not dimensionally reduce to all orders in the gauge coupling
expansions. An alternative approach of effective model construction
was proposed by Nadkarni\cite{nad}. Following this approach, an effective
 static model in three dimensions
 was proposed reasonable describing the full four dimensional
theory at the distance range $RT \cong 1$ at sufficiently high
temperature in the deconfinement phase\cite{reis}. Moreover, it was
confirmed at a nonperturbative level, that the dimensional reduction
gives reasonable results beyond the perturbative horizon\cite{cock}.

Inspired by the dimensional reduction calculations
in continuum $SU(2)$ gluodynamics at finite temperatures, we obtain
in a model independent way reasonable
results for heavy quarks potential. The resulting potential has
expected asymptotic behaviour in confinement
and deconfinement phases too. Moreover, we found an evidence of a phase
transition corresponding to the restoration of $SU(2)$ symmetry at high
temperatures from Abelian gauge symmetry at lower temperatures in the three
dimensional static model. In Sec.2 we briefly discuss the general picture
of a finite temperature theory and the problem of the effective potential.
The method for the calculation of the heavy quark potential is described in
Sec.3. The results are summarized in Sec.4. Sec.5 contains brief discussion.
2. Let us briefly discuss the general picture encountered in the
finite temperature theory. The formalism was introduced by
C. Bernard\cite{bern} by the partition function of the theory
in the functional integration formalism:

\begin{equation}
Z(\beta)=N\int [DA_{\mu}^a]\ e^{-S_E},
\end{equation}
\noindent
where $S_E$ means Euclidean action obtained by Wick rotation from $SU(2)$
gauge field action and given by the relation

\begin{equation}
S_E = \frac{1}{4}\int_{0}^{\beta}dt \int d^3 x F^a_{\mu \nu}F^a_{\mu \nu}.
\end{equation}
\noindent
The gauge fields are periodic in the Euclidean time variable

$$
A_{\mu}^a(t+\beta,{\bf x}) = A_{\mu}^a(t,{\bf x}),
$$

\noindent
(or antiperiodic in the case of fermionic fields).
The length of periodicity interval $\beta$ can be identified
with the inverse temperature following
the similarity of the finite temperature formalism with the formalism
of the statistical physics.
Using Fourier transform the Euclidean time integration is replaced
by the infinite summation over Fourier modes of the gauge fields.

One of the methods for the study of the phase structure in the gauge
fields theory rely on the conjecture that the infrared sector of
the finite temperature theory possesses the same phase structure
as the complete field theory. Due to Fourier transform the nonzero
Fourier modes of the gauge fields acquire the temperature dependent mass
terms. The only candidate for dynamical variables of the infrared sector
of the theory are therefore the Fourier zero modes of fields.
To obtain an effective theory for such modes we must integrate out all
non-static modes. Corresponding functional
integration was subjected to different approximations, one of them is
discussed in detail in our previous article\cite{ja}. The result is
static, 3-dimensional effective theory for zero Fourier
modes of the gauge fields.
The effective action can be written in the form:

\begin{equation}
S_{eff} = \beta \int d^3x [\frac{1}{4}F^a_{ij}F^a_{ij} +
\frac{1}{2}(D_i {\cal A}^{a}_{0})^2] + V_{eff}
\label{efac}
\end{equation}

\noindent
where
$$
F^a_{ij}= \partial_i {\cal A}^a_j - \partial_j {\cal A}^a_i
-g\varepsilon^{abc}{\cal A}^b_i {\cal A}^c_j
$$
\noindent
depends on the static modes ${\cal A}^a_i$
of the gauge potentials. The effective potential $V_{eff}$ is the rather
complicated functional\cite{ja} of the static fields resulting from the
functional
integration over nonstatic Fourier modes. Let us to remark that the system
described by Eq.(\ref{efac}) can be related to the Higgs - Yang - Mills
4-dimensional system described by the Lagrangian density:

\begin{equation}
L = -\frac{1}{4}G^a_{\mu \nu}G^{a\mu \nu} -\frac{1}{2}
D_{\mu}\Phi^aD^{\mu}\Phi^a - V(\Phi)
\label{hym}
\end{equation}

\noindent
In the temporal gauge $A_0^a=const$ the static solution of the Eq.(\ref{hym})
corresponds to the system of Eq.(\ref{efac}), if we identify\cite{polo} the
Higgs field with the chromoelectric field ($\Phi^a = {\cal A}_0^a$) and
denote $F_{ij}^a=G_{ij}^a$.

In our approach,
the effective potential in the Eq.(\ref{efac})	is calculated
by Gauss integration around the classical minimum $A^a_i=0$ and
$A^a_0=const$, $a=1,2,3$
of the pure gauge action. There are two possibilities:

1) The classical minimum of the global $SU(2)$ action is achieved for the
values $A_0^a = 0, \ a=1,2,3,$ which simultaneously correspond to the minimum
of effective potential $V_{eff}$ in Eq.(\ref{efac}). The
resulting effective system
is in the symmetry unbroken state and it describes
the interaction of the massive Higgs particle ${\cal A}^a_0$ with the massless
gauge bosons ${\cal A}^a_i$. This picture was confirmed by analytical
calculations\cite{ja,gock,ole}.

2) The classical minimum of the global $SU(2)$ action is achieved if some of
$A_0^a \ne 0$, and as a result the minimum of the effective potential
changes its position too. The whole physical picture significantly changes
and we can follow the general scheme of the Higgs mechanism.

In the second case we proceed as follows.
Using a suitable unitary gauge
transformation\cite{huang} $({\cal A}^a_i \rightarrow B^a_i,\ {\cal A}^a_0
\rightarrow C^a_0)$ we can separate the chromoelectric fields into one massive
mode (choosing $A^3_0 = C_0 + a^3_0$, $C_0=const.$) and two "would be"
Goldstone
modes which disappear after unitary gauge transformation:

\begin{eqnarray}
S_{eff} = \beta \int d^3x [\frac{1}{4}G^a_{ij}G^a_{ij} +
\frac{1}{2}(gC_0)^2[(B^1_i)^2 + (B^2_i)^2] +
\frac{1}{2}(\partial_i a^{3}_{0})^2]  + V_{eff} \\
G^a_{ij}= \partial_i B^a_j - \partial_j B^a_i
-g\varepsilon^{abc}B^b_i B^c_j.    \nonumber
%\label{efac1}
\end{eqnarray}

\noindent
$C_0$ is the background field value where the global $SU(2)$ action
and the effective potential in Eq.(\ref{efac}) possesses the minima.
Two chromomagnetic fields modes (following our choice  $B^1_i$ and
$B^2_i$) become massive with the mass terms proportional to $|C_0|$.
The massless chromomagnetic field in tree approximation has the same colour
signature as the persisting chromoelectric field. Two chromomagnetic massive
fields
modes now satisfy the conditions of the Appelquist - Carazzone decoupling
theorem\cite{cara,appis}. We integrate them out to obtain the infrared sector
of the original theory. In the tree approximation level this effective theory
possesses $U(1)$ symmetry. The behaviour beyond tree level depend on the
effective potential. Calculations up to second order in the chromomagnetic
potentials show interesting features\cite{ja}.
The functional integration over the massive
fields introduces into effective action the contributions
of the form:
$$
\sum_{-\infty}^{\infty} \frac{1}{{|n+X|}^{\alpha}},\ \ \
\alpha =const.,
$$
\noindent
where we introduced instead of the chromoelectric field the new variable
$$
X=\frac{g \beta C_0}{2 \pi}.
$$

\noindent
This lead to the important
conclusion that the effective potential is even and periodic in
the nonzero background value of the chromoelectric field $C_0$:

\begin{eqnarray}
V_{eff}(-X)\ =\ V_{eff}(X), \nonumber	     \\
V_{eff}(X)\ =\ V_{eff}(X+1),
\label{symm}
\end{eqnarray}
\noindent
where $g$ is coupling constant,
$\beta$ is the periodicity interval in the imaginary time.
The partition
function  can be expressed as a sum of functional integrals defined on disjoint
intervals of values of the field $A^3_0$ of unit length but, due to the
periodicity, with the same action on each interval.
Moreover, the Polyakov loop operator is the same on different intervals
in questions. If we calculate the mean value of an operator,
which can be rewritten as a function of the Polyakov loop instead of  $A^3_0$,
it is sufficient to calculate the corresponding functional integral
over one interval of the periodicity of the effective action. The mean value
of the field $A^3_0$ within one such interval can have a nonzero value.
Nevertheless, integrating over the whole functional space of $A^3_0$,
the individual nonzero values cancel each other due to the symmetry
Eq.(\ref{symm}) of the action functional. The resulting mean value of the field
$A^3_0$ is therefore zero in accordance with the Elitzur theorem\cite{eli}.

The behaviour of the $V_{eff}(X)$ for $X\in (0,1)$ for different values of
the coupling constant was studied. The results for $\frac{\partial V_{eff}}
{\partial X}$  are summarized
in Fig.1. To clarify its meaning  we present in Fig.2
the possible shapes of the corresponding effective potentials.
The identification of the different phase transitions is obvious.
We do not study in
detail the different phases and the corresponding phase transitions in this
article. Our goal is to show that the phase structure picture is
not in contradiction with the accepted finite temperature $SU(2)$ system
behaviour known from the lattice calculations. For this aim we calculate
the  heavy quark potential as a temperature dependent quantity in the
continuum $SU(2)$ system with (infinitely) heavy quarks.

3. The heavy quark potential $V_{qq}$ in the finite temperature theory is
connected with the Polyakov loop correlator\cite{jaf} as follows
\begin{equation}
{\rm e}^{-\beta V_{qq}({\bf x_{1}},{\bf x_{2}})}= \langle L({\bf x_{1}}),L({\bf
x_{2}}) \rangle.
\label{bas}
\end{equation}
\noindent
In the static gauge the Polyakov loop  is given by the following
expression
\begin{equation}
L({\bf x}) = \cos [\frac{1}{2}\beta g (C_0 + a^3_0({\bf x}))],
\end{equation}
\noindent
i.e. Polyakov loop depends only on the chromoelectric field. This
justifies us to use for the calculation of the correlator the same method as we
used for the calculation of the effective potential. Since the evaluation of
the
effective potential has been performed in detail in Ref.(\cite{ja}) we
sketch here only the main steps in the calculation. For the partition
function as a functional of the effective action with the fields $a_0^3,
a_i^3$ as the dynamical variables, was obtained the expression:

\begin{equation}
Z(\beta ) = \int [Da^3_0][Da^3_i] {\rm e}^{-S_{eff}},
\label{partf}
\end{equation}
\noindent
the effective action is given as
\begin{eqnarray}
S_{eff}= \frac{1}{\beta}\int d^3x[(\partial_ia^3_0)^2 + (\partial_jB^3_k -
\partial_kB^3_j)^2] + V_{eff}(X,a^3_0,B^3_i), \nonumber \\
V_{eff}(X,a^3_0,B^3_i)=- \frac{V}{\beta ^3} \left[\frac{3\pi^2}{45}-
\frac{4\pi ^2}{3} (X^4-2X^3+X^2)\right] + \nonumber \\
\int \frac{d^3\kappa}{(2\pi)^3}
\tilde{a}^3_0({\bf {\kappa}}) S^{eff}_{cor}(\kappa,X) \tilde{a}^3_0(-{\bf
{\kappa}}).
\label{pot}
\end{eqnarray}

\noindent
In the last expression $\tilde{a}^3_0({\bf {\kappa}})$ is the Fourier
transform of the function $\beta a^3_0 ({\bf x})$, $\kappa=\beta k$ is the
dimensionless momentum, $V$ is the volume of the 3-dimensional space.
 $S^{eff}_{cor}(\kappa,X)$ was calculated in Ref.(\cite{ja})
in the approximation where the functional integration over the second order
in the chromomagnetic gauge field was performed. This calculation gives
 the relation:

\begin{eqnarray}
\lefteqn{S^{eff}_{cor}(\kappa,X) =}		       \nonumber \\
&&  g_r^2 \left\{ 2(X^2-X+\frac{1}{6}) -
\frac{11}{3} \hat{\kappa}^2[(1-\ln(4\pi )-\Psi (X)-\Psi(1-X)]
+\frac{(\hat{\kappa})^3\pi^3}{\sin^2(\pi X)}  \right.  \nonumber \\
&& + \sum_{-\infty}^{\infty} \left[-\left(
\mid n+X\mid^2 + \frac{2\hat{\kappa}^4}{\mid n+X\mid^2}
+2\hat{\kappa}^2 \right)
\frac{1}{\hat{\kappa} }
\arctan\frac{\hat{\kappa} }{\mid n+X\mid}  \right.  \nonumber  \\
&& \left. \left. + \mid n+X\mid +
\frac{5}{3}\frac{\hat{\kappa}^2}{\mid n+X\mid }
\right] \right\}
\label{pote}
\end{eqnarray}
\noindent
where
$$
\hat{\kappa}= \frac{\kappa}{4\pi}
$$
\noindent
Inserting this relation into Eq.(\ref{pot}) for the effective action,
we can see that first term of Eq.(\ref{pote}) generate the $(mass)^2$ term
for the quantum corrections field $a_0^3(\bf x)$ and the rest of this relation
could introduce the higher derivatives of such field. The problem
encountered in this calculation is negative value of the $(mass)^2$
for some interval of the values of $X$. We have an indication that this is
caused by some abandoned terms of the higher order
in the chromomagnetic fields which contribute through functional integration
to the second order term in the quantum correction field $a_0^3(\bf x)$.
However, because this long complicated calculation are not yet finished,
we could estimate its possible effect by adding a small positive constant
into $S^{eff}_{cor}(\kappa,X)$ modifying it into positive definite quantity
$S^{mod}_{cor}(\kappa,X)$.

We calculated the Polyakov loop correlator in the same approximation as the
effective potential in Ref.(\cite{ja}) and we obtained the expression:

\begin{equation}
\langle L(0),L(R) \rangle = \frac{1}{2}{\rm e}^{-\frac{g^2_r}{8} S^{-1}(0,0)}
[{\rm e}^{\frac{g^2_r}{8} S^{-1}(0,R)} + \cos (2\pi X)
{\rm e}^{-\frac{g^2_r}{8} S^{-1}(0,R)}].
\label{inv}
\end{equation}
\noindent
This relation is the most important result of this brief report. It permits
us to calculate the heavy quark potential as a function of the distance and
the temperature (by means of temperature dependent renormalized coupling
constant $g_r$). In the last equation $S^{-1}({\bf x},{\bf y})$ means the
inverse matrix element (with continuum indices {\bf x} and {\bf y}) given
by the relation\cite{vladi}:

\begin{equation}
S^{-1}({\bf x},{\bf y}) = \int \frac{d^3 (\beta k)}{(2\pi )^3}
\frac{1}{S^{mod}_{cor}(k,X)} {\rm e}^{-i{\bf k}({\bf x} - {\bf y})}
\label{inve}
\end{equation}

\noindent
We evaluated the integral in Eq.(\ref{inve}) in polar
coordinates. The radial integration can be rewritten in the form:

\begin{equation}
S^{-1}(0,R) = \frac{1}{i(2\pi )^2\rho } \{ \int_{-\infty}^{0}
\frac{\kappa\ d\kappa}{S_{cor}^{mod}(-\kappa,X)}\ {\rm e}^{i\kappa \rho}
+ \int_{0}^{\infty }
\frac{\kappa\ d\kappa}{S_{cor}^{mod}(\kappa,X)}\ {\rm e}^{i\kappa \rho} \}
\label{inv2}
\end{equation}
\noindent
where $\rho = T R$.
The singular structure of the integrand in the complex $\kappa$ plane is
defined
by the singular structure of the $S^{mod}_{cor}$.
The function $S^{mod}_{cor}(\kappa,X)$
possesses two cuts, $(-i\infty, -iX), (iX, i\infty)$ in the complex $\kappa$
 plane. Because the integrands in Eq.(\ref{inv2}) have no singularities
outside of real and imaginary axes and moreover,
\begin{equation}
\lim_{\mid \kappa \mid \rightarrow \infty} \kappa \left[S_{cor}^{mod}(\kappa
,X)
\right]^{-1} =0, \nonumber
\end{equation}
\noindent
we can change the range of the integration by closing properly the integration
contours. Then, we can rewrite the Eq.(\ref{inv2}) in the form:
\begin{equation}
S^{-1}(0,R) = \frac{1}{i(2\pi )^2\rho } \left\{- \int^{i\infty}_{i0}
\frac{\kappa\ d\kappa}{S_{cor}^{mod}(-\kappa,X)}\ e^{i\kappa \rho}
+ \int_{i0}^{i\infty }
\frac{\kappa\ d\kappa}{S_{cor}^{mod}(\kappa,X)}\ e^{i\kappa \rho} \right\}
\nonumber
\end{equation}
\noindent
By substitution  $\kappa = i\tau$ we obtain:
\begin{equation}
S^{-1}(0,R) = \frac{2}{(2\pi )^2\rho }\int_{0}^{\infty} d\tau e^{-\tau \rho}
\ \frac{\tau\ Im S^{mod}_{cor}(i\tau, X)}{\mid S^{mod}_{cor}(i\tau, X) \mid^2}
\label{fin}
\end{equation}
Applying the mean value theorem to the integral in
Eq.(\ref{fin}) we can  estimate $S^{-1}(0,R)$ as
\begin{equation}
S^{-1}(0,R) = C(T)\ \frac{{\rm e}^{-\mu TR}}{TR},
\label{est}
\end{equation}
\noindent
where $C(T)$ is a temperature dependent quantity, and the $\mu$ is a
positive constant. In principle $C(T)$ is calculable and the quantity $\mu$
can be obtained from the Fourier transform of $S^{eff}_{cor}$.
 In the lowest approximation $\mu{\beta}^{-1}$ is the mass term of the quantum
fluctuations of the field $a^3_0$. Since we modified $S^{eff}_{cor}$ to
$S^{mod}_{cor}$ by an {\it ad hoc} constant, such calculations are meaningless
at the moment, and we treat
Eq.(\ref{est}) qualitatively. Eq.(\ref{est}) is valid for $R\neq 0$. The matrix
element $\frac{g_r^2}{8\beta}S^{-1}(0,0)$ is infinite and represents the quark
self-energy $V_{qq}^{self}$ in the
heavy quark potential given in Eq.(\ref{bas}). In what follows we shall
consider
only the interaction energy of the quarks by subtracting this self-energy:

$$
V_{qq}^{int} = V_{qq} - V_{qq}^{self}
$$
The factor containing $S^{-1}(0,0)$ then drops out from all relations and
we have

\begin{equation}
{\rm e}^{-\beta V_{qq}^{int}}
 = \frac{1}{2} [{\rm e}^{\frac{g^2_r}{8} S^{-1}(0,R)} + \cos (2\pi X)
{\rm e}^{-\frac{g^2_r}{8} S^{-1}(0,R)}].
\label{inv1}
\end{equation}

We should
 like to stress the importance of the Eq.(\ref{est}). All qualitative features
of the discussion what follows are based on the particular form
of r.h.s. of Eq.(\ref{est}) and do not depend on details of $S_{cor}^{eff}$ in
Eq.(\ref{pot}). Our result now simply follow from Eqs.(\ref{est},\ref{inv1})
and they are summarized bellow.

4. We find that for the coupling
constant values $g_{r} \geq 5.5$ (in accordance with the previous work\cite{ja}
) the effective potential has the global minimum at $X = 1/2$. To this
value of $X$ corresponds the vanishing value $\langle L \rangle = 0$ of the
Polyakov loop. This is characteristic for the confinement phase.
The heavy quark potential
calculated from the Eqs.(\ref{bas},\ref{est},\ref{inv1}) possesses the desired
$R$ dependence because the leading term for $R \rightarrow \infty$ is linear:
\begin{equation}
\frac{1}{T} V_{qq}^{int}(R) = \mu RT + const + \ln{RT} + o(R^{-2})
\end{equation}

We find the two substructures
in the non-confinement phase characterized by $\langle~L~\rangle~\neq~0$.
In the first case, connected to the
confinement phase by the second order phase transition, the effective potential
possesses two minima for $X \neq 1/2$ symmetric with respect to $X = 1/2$.
The heavy quark potential in the non-confinement phase is usually calculated
according to the formula\cite{irb}:

\begin{equation}
\frac{1}{T}V_{qq}(R) =- \ln{\frac{\langle L(0),L(R)\rangle }{(\langle L(0)
\rangle)^2 }}
\label{decpot}
\end{equation}
\noindent
Inserting Eqs.(\ref{est},\ref{inv}) into the last equation we find for
$R \rightarrow \infty$ the result:
\begin{equation}
\frac{1}{T}V_{qq}(R) = {\tan }^2(\pi X)\ \frac{g_r^2}{8}\ C(T)\ \frac
{{\rm e}^{-\mu TR}}{TR} \ + \ o(R^{-2}).
\label{debye}
\end{equation}
\noindent
In the leading term of the last expression we recognize the Debye screening of
the colour charge potential in accordance with the lattice results\cite{irb},
as well as with the analytical calculations\cite{zin}. Our results are also
consistent with the predictions of Svetitsky and Yaffe\cite{jaf}.

Interesting structure appears for higher temperatures corresponding
to smaller values of the coupling constant $g_r$. If the value of $g_r$
is decreasing, the minimum of $V_{eff}$ at $X_0 \in (0,\ 1/2)$ jumps
to the value $X_0=0$ (and the minimum at the point $X_1=1-X_0$ jumps to
$X_1=1$). The effective potential
as the function of the coupling constant behaves like a classical potential
for the first order phase transition. We deduce that this phase transition
takes
place at $g_{cr2}\approx 3.$ The value
of the Polyakov loop in the new phase is $\langle L \rangle = \pm 1$. The
heavy quark potential is given in the high temperature region according to
Eq.(\ref{decpot}) by the next to leading term in Eq.(\ref{debye}):

\begin{equation}
\frac{1}{T}V_{qq}(R)=
\frac{1}{2}\ \left( \frac{g_r^2}{8}\right) \ C^2(T)\ \frac{{\rm e}^{-2\mu TR}}
{(TR)^2}\ +\ o(R^{-4}).
\end{equation}
\noindent
The leading term of the last expression possesses the same $R$ dependence
as the heavy quarks potential for very high temperatures calculated
perturbatively. Such behaviour of the heavy quark potential at very high
temperatures as we find here was reported by Gao\cite{gao} in the
framework of lattice calculations. We can estimate the critical temperature by
the running coupling constant relation\cite{satz}
\begin{equation}
g_r^2(T) = \frac{24 \pi^2}{22 \ln{\frac{T}{\Lambda}}},
\end{equation}
\noindent
here $\Lambda$ is the QCD cut-off parameter.
For the critical coupling constant $g_{cr2}$ we estimate the critical value
of the temperature $T_{cr2}\approx 3.3\Lambda$. This means that
$T_{cr2}\approx 2.2T_{cr}$, where $T_{cr}$ is the temperature of the phase
transition where the value of the Polyakov loop becomes nonvanishing.

We conjecture that this first order phase transition is connected with
restoration of the $SU(2)$ gauge symmetry for the static system at high
temperatures. The proof could be based on the more precise calculation of the
effective potential.  We conjecture that together
with the jump to the minimum of the potential at $X=0$, corresponding to
the classical value of the chromoelectric field
$A_0=0$, the system in question
undergoes a phase transition to deconfined $SU(2)$ symmetric phase described by
the effective action with three independent massless chromomagnetic static
fields and chromoelectric field identified as the Higgs field.

5. To conclude, we presented here some arguments for the support of more
detailed study of the effective systems in the continuum finite temperature
theory. The results of the heavy quark potential investigations presented here
indicate that the dimensional reduction Ansatz in the continuum integral method
goes beyond the perturbative horizon. We find a qualitative evidence for
possible phase transition corresponding to restoration of the $SU(2)$ symmetry
at very high temperature.  In the framework of the calculation proposed in
ref.\cite{ja,va} this phase transition is of the first order but the final
conclusion could be presented only after more detailed investigations.

\begin{figure}
\caption{The derivative of the effective potential $\frac{\partial V_{eff}}
{\partial X}$ for the different values of
the coupling constant $g_r$. The variable $X$ is defined in the text.
The derivative is symmetric function on the interval $(0,1)$.The value
$g_r=0$ corresponds to the full line; $g_r=2$ correspond to the dashed
line; $g_r=4$ correspond to the dash-dotted line and $g_r=6$ correspond
to the dash-dot-dotted line. The analytic calculations were checked and the
numeric ones were performed by the program
MATHEMATICA\protect\cite{wolf}.}
\label{fig1}
\end{figure}

\begin{figure}
\caption{The effective potentials corresponding to the
behaviour of the different derivative described in
Fig.1. Case a) correspond to $g_r=0$, for b) we chose
the first order phase transition at $g_r \approx g_{cr2}$. Case c)
corresponds to the effective potential in deconfinement phase where
$g_{cr2} < g_r < g_{cr}$. The derivative for $g_r>g_{cr}$
belongs to the class d) displayed in Fig.1 and it is characteristic for the
confinement phase.}
\label{fig2}
\end{figure}
\end{document}